\documentclass[fleqn,usenatbib]{mnras}

\usepackage{newtxtext,newtxmath}
\usepackage{graphicx}
\usepackage{array}
\usepackage{hyperref}

\usepackage[T1]{fontenc}
\usepackage{ae,aecompl}
\usepackage{rotating} 
\usepackage{pdflscape}

\usepackage{graphicx}	
\usepackage{amsmath}	
\usepackage{amssymb}	

\title[Possible detection of a new cyclotron feature in 4U 1700-37]{Possible detection of a new cyclotron feature in 4U 1700-37}

\author[Bala et al.]{
Suman Bala\thanks{E-mail: bala@iucaa.in},
Jayashree Roy\thanks{jayashree@iucaa.in} and
Dipankar Bhattacharya
\\
Inter University Center for Astronomy and Astrophysics, Ganeshkhind, Post Bag 4, Pune 411007, India.\\
}

\date{Accepted XXX. Received YYY; in original form ZZZ}

\pubyear{2015}

\begin{document}
\label{firstpage}
\pagerange{\pageref{firstpage}--\pageref{lastpage}}
\maketitle

\begin{abstract}
We present a spectral and timing study of the High Mass X-ray Binary transient source 4U 1700-37 using NuSTAR and ASTROSAT/LAXPC. The source is observed in two different flux states. A combined spectral analysis of NuSTAR's FPMA and FPMB shows the possible hint of a cyclotron line feature at $\sim$16 keV. The line feature is consistently present in different continuum models with at least 3$\sigma$ confidence level. We do not detect the presence of a previously reported ~39 keV cyclotron line in the combined spectra. A $\sim$16 keV cyclotron feature would suggest that the compact object is a neutron star with a magnetic field strength $\sim2.1\times$10$^{12}$ Gauss in the emission region. We also find the presence of a rare Ni $K{\alpha}$ emission line around 7.6 keV in the NuSTAR spectrum. We searched the NuSTAR and ASTROSAT data for coherent or quasi-periodic oscillation signals but found no evidence in the frequency range 0.1 mHz to 10$^3$ Hz.  
\end{abstract}




\section{Introduction}
The X-ray source 4U 1700-37 was first detected by the Uhuru satellite in 1970 December \citep{1973ApJ...184L..65J}. This High Mass X-ray Binary (HMXB) has an orbital period of 3.41 days. The extreme 07 f star HD 153919 is a confirmed optical counterpart \citep{1973MNRAS.163P...7P, 1973MNRAS.163P..13H, 1980ApJ...238..238D}, of which an orbital eccentricity of 0.22$\pm$0.04 and a radial velocity semi-amplitude of 20.6$\pm$1.0 km s$^{-1}$ have been measured \citep{2003A&A...407..685H}. The binary parameters of the 4U 1700-37 System are tabulated in \citet{2016ApJ...821...23S} and \citet{2016MNRAS.461..816I}.\par

One of the primary diagnostics of the presence of a neutron star (NS) in an X-ray binary system is the detection of pulsation. A persistent, approximately sinusoidal modulation of $\sim$60\% amplitude, and 97 min periodicity was reported from the source 4U 1700-37 \citep{1978ApJ...224L.119M}. \citet{1978IBVS.1424....1K} further examined the optical data of HD 153919 and found a similar $\sim$97 min periodicity in 0.44-0.59 binary orbital phase. Another 96.8-min periodicity was reported by \citet{1978ApJ...224L.119M} but later \citet{1979ApJ...228L..75H} showed that it was caused by instrumental effects . The 96.8 min periodicity was also rejected by 
\citet{1980ApJ...238..238D} studying OSO 8 data above 20 keV. \citet{1984PASJ...36..691M} detected a 67.4s periodicity with a $\sim$4 
percent modulation of the X-ray flux during a single 20-min flare, observed by the X-ray satellite Tenma. \citet{1986MNRAS.222P..21G} reported 
the absence of any coherent periodicity over the frequency range $2\times10^{-5}$ to 256 Hz. Similarly, no periodicity was found in the 
timescales of 32 ms to 2 hr from EXOSAT observations \citep{1987A&A...173...86D}. \citet{1996ApJ...463..297B} suggested that
the source may be a black hole candidate based on the absence of pulsations and the presence of a hard X-ray tail in the spectrum.
The power density spectrum (PDS) of the source was reported to show a 6.5 mHz QPO \citep{2003ApJ...592..516B} with a fractional root mean square (rms) amplitude of 4.5$\%$ before the eclipse. Recently, \citet{2015MNRAS.448..620J} strongly suggested the absence of any coherent periodicity, and reported a $\sim$20 mHz QPO from SUZAKU HXD/PIN data.\par

The source 4U 1700-37 is known to be a wind fed accreting HMXB system \citep{1989ApJ...343..409H}. The stellar wind absorbs the low energy X-ray photons by
photoelectric absorption which is found to be orbital phase dependent. \citet{1989ApJ...343..409H} studied the variation of the 
X-ray absorption in different orbital phases, using a continuous observation of $\sim$3.2 days by the EXOSAT satellite. They 
observed a sharp increase in absorption near phase 0.6 to eclipse ingress and a decrease in absorption from phase
0.5 to eclipse egress. The hydrogen column density ($N_H$) is found to vary between 1$\times$10$^{22}$ and 45$\times$10$^{22}$  cm$^{-2}$
during this 3.2 day-long observation. 
\citet{1994A&A...289..846K} studied the optical lines in the spectra of optical counterparts of Vela X-1 and 4U 1700-37. They found an increment in blue-shifted absorption before and during the transit of the compact object through the line of sight towards the 
companion star. They argued that the observed obscuration or eclipse of the companion 
star in Vela X-1 could not be explained by an accretion wake surrounding the compact object or a gas stream resulting from tidal forces. They attributed the absorption seen in the spectrum to a photo-ionization wake trailing the X-ray source. Changes in the geometry of the photo-ionization wake could account for the orbit wise variation of the absorption feature.\par

Another clear diagnostic for the presence of a NS in an X-ray binary system is a cyclotron line. Cyclotron lines, also called cyclotron resonant scattering features (CRSFs) are absorption features in the X-ray spectrum of highly magnetized neutron stars with teragauss magnetic field ($B$). The cyclotron line energy ($E_{cyc}$)  allows a direct measurement of the magnetic field strength in X-ray Pulsars,
\begin{equation*}
 E_{cyc}=11.6\times B_{12}(1+z)^{-1} keV
\end{equation*}
Where $B_{12}$ is the magnetic field in the unit of $10^{12}$ Gauss and $z$ is the gravitional redshift from the line formation region near the NS.
Till now, nearly 36 sources are known to exhibit CRSF in their x-ray spectrum \citep{2019A&A.staubert..622A..61S}. In a few cases, multiple absorption features have also been observed and identified as harmonics  (e.g. 4U 0115+63: \citet{1999ApJ...521L..49H}, Cep X-4: \citet{2015MNRAS.453L..21J}). But in several sources, e.g., 4U 1538-52
\citep{2009A&A...508..395R}, Vela X-1 \citep{2002A&A...395..129K}, departures from an integral
line ratio have been observed. Most of the observed line features
are consistent with a simple Gaussian shape, primarily due to the limited
spectral resolution of the observing instruments. In very few cases,
asymmetric or distorted line profiles have also been observed, like
Cep X-4; \citep{2015ApJ...806L..24F}. The CRSF line energy is also found to vary with spin phase (e.g. Vela X-1; \citet{2013ApJ...763...79M}, A 0535+26; \citet{2013ApJ...771...96M}, GX 301-2: \citet{2004A&A...427..975K}). In a few sources, dependence of the CRSF on X-ray luminosity has also been observed (4U 0115+63 \citet{2006AdSpR..38.2756N}; A0535+262 \citet{2006ApJ...648L.139T}) . The correlation is found to be either positive 
(e.g.\ Her X-1: \citet{2007ESASP.622..465S}; GX 304-1: \citet{2011PASJ...63S.751Y};
Cep X-4: \citet{2015ApJ...806L..24F}) or negative (e.g.\ V 0332+53; \citet{2017MNRAS.466.2143D}). Long term variation of the energy of the cyclotron line has also been observed in 
two sources, namely  Her X-1 \citep{2019MNRAS.Ji.484.3797J,2017A&A...606L..13S}, and Vela X-1 
\citep{2014ApJ...780..133F,2019MNRAS.Ji.484.3797J}. \citet{1999A&A...349..873R} observed the first 
broad-band spectrum of 4U 1700-37 in the energy range 0.5-200 keV using BeppoSAX. The spectrum was well described by an absorbed 
powerlaw with a cutoff, typical of X-ray binary pulsars. The spectrum showed a soft bremsstrahlung component at 
$\sim$2 keV. They suggested the possibility of a cyclotron line at $\sim$37 keV, which was a direct evidence for the compact object being a NS. \citet{2015MNRAS.448..620J} further endorsed 
the findings of \citet{1999A&A...349..873R} with a detection of a broad (width 19$^{+6.1}_{-4.3}$ keV) cyclotron line 
like feature at $\sim$39 keV in the broadband spectrum derived from a SUZAKU observation of September 13-14, 2006. This detection implied a
surface magnetic field $\sim$3.4$\times$10$^{12}$ Gauss of the neutron star.
In a recent work, \citet{2016ApJ...821...23S} described the source spectrum with a 
more physical model, and found no evidence of any cyclotron line near 40 keV in the BeppoSAX and SUZAKU Data. They studied the high-soft state spectral 
evolution of the source by modelling the spectra with two comptonization components. They observed that the componized spectral components are similar to those previously found in NS systems and concluded that source 4U 1700-37 is a NS
binary system.\par

This paper is structured as follows: introduction in section 1 is followed by a description of observations and data analysis in section 2. In section 3, we discuss the results derived from timing and spectral analysis. All the errors are reported in this paper with 90$\%$ confidence level. 

 \begin{figure*}
\centering
\includegraphics[scale=.3]{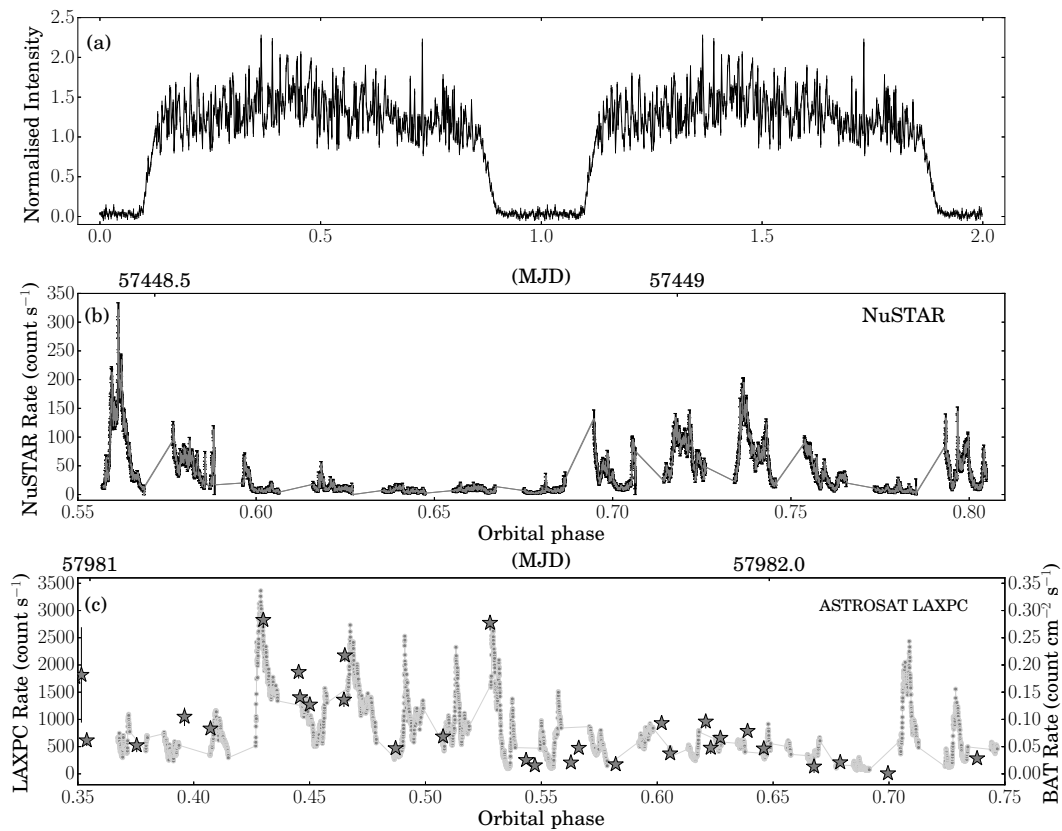}
\caption{a) The SWIFT/BAT light-curve of 4U 1700-37 during MJD 57000-58500 folded with a reference epoch of 57446.55 MJD. b) The 3.0-80.0 keV lightcurve of NuSTAR observation. c) The 15-50 keV LAXPC10 lightcurve (light grey points) over plotted with SWIFT-BAT lightcurve (dark grey Stars). The data gaps in both light-curves correspond to source occultation and passage through the South Atlantic Anomaly (SAA) region. }
\label{fig:all_lc}
\end{figure*}

\section{observations and data analysis}
In our analysis, we have used two archived data sets. The first is a 40 ks NuSTAR observation (Obs Id: 30101027002, PI. Felix Fuerst)\footnote{\hyperlink{https://heasarc.nasa.gov/cgi-bin/W3Browse/w3browse.pl}{https://heasarc.nasa.gov/cgi-bin/W3Browse/w3browse.pl}} obtained on 
1st March 2016. The second 
 dataset is from an ASTROSAT LAXPC observation carried out on 16th Aug 2017 (Obs Id:9000001464, PI: Gaurava K. 
Jaisawal)\footnote{\hyperlink{https://astrobrowse.issdc.gov.in/astro\_archive/archive/Home.jsp}{https://astrobrowse.issdc.gov.in/astro\_archive/archive/Home.jsp}}. \par

\begin{table}
\caption{Observation log.}
\centering
\begin{tabular}{p{1.5cm}p{1.5cm}p{0.75cm}p{1.25cm}p{1.25cm}}
\hline
Instrument & Obs Date & MJD & Stare time & Orb. Phase \\
\hline \hline

NuSTAR & 01/03/2016 & 57448 & 73.08 ks & 0.55-0.80 \\
\hline
ASTROSAT (LAXPC) & 16/08/2017 & 57981 & 131.80 ks & 0.35-0.75 \\
\hline
\end{tabular}
\end{table}

NuSTAR consists of two identical Focal Plane Modules, A and B (FPMA and FPMB, respectively) \citep{2013ApJ...770..103H}. We have extracted data separately from both the modules. We have used \texttt{nupipeline} version 0.4.6 distributed with HEASOFT\footnote{\href{https://heasarc.gsfc.nasa.gov/docs/software/heasoft/}{https://heasarc.gsfc.nasa.gov/docs/software/heasoft/}} v6.25,
and calibration files 20181030, to create clean event files. We observe that the SNR (signal to noise ratio) at 3--79 keV increases with the radius of the source region, and levels off beyond $\sim$90 arc--second. The 40--79 keV SNR is found to increase with radius and reach a maximum at 90", beyond which it starts to decrease. So, we have selected the source region as a 90" circle around the brightest point and the background is obtained from a same size region away from the source. We have then used \texttt{nuproducts} to create the spectra and light-curves. \par

 ASTROSAT is the first Indian multi-wavelength space observatory consisting of five principal scientific payloads \citep{2006AdSpR..38.2989A}.
 ASTROSAT can observe in different energy ranges: visible (320-530 nm), near Ultra Violet (NUV: 180-300 nm), far UV (FUV: 130-180 nm), 
 soft X-rays (Soft  X-ray  imaging  Telescope (SXT): 0.3-8.0 keV), Scanning Sky Monitors  ( SSM: 2.0-10.0 keV) and hard 
 X-rays (Large Area X-ray Proportional Counters (LAXPC): 3-80 keV, Cadmium-Zinc-Telluride Imager (CZTI: 20-150 keV). \par

 We have used publicly available LAXPC data for the timing analysis of the source. The standard available software \texttt{LAXPCsoftware} (Format (A)) from 
 ASTROSAT Science Support Cell (ASSC\footnote{\href{http://astrosat-ssc.iucaa.in/}{http://astrosat-ssc.iucaa.in/}}) has been used to
 create the event-file, good time intervals (GTI) file, light-curves, and background light-curves. The GTI is further adjusted to remove unusual high or low counts before generating the final lightcurve.
 
 \begin{figure*}
\centering
\includegraphics[width=16.5cm]{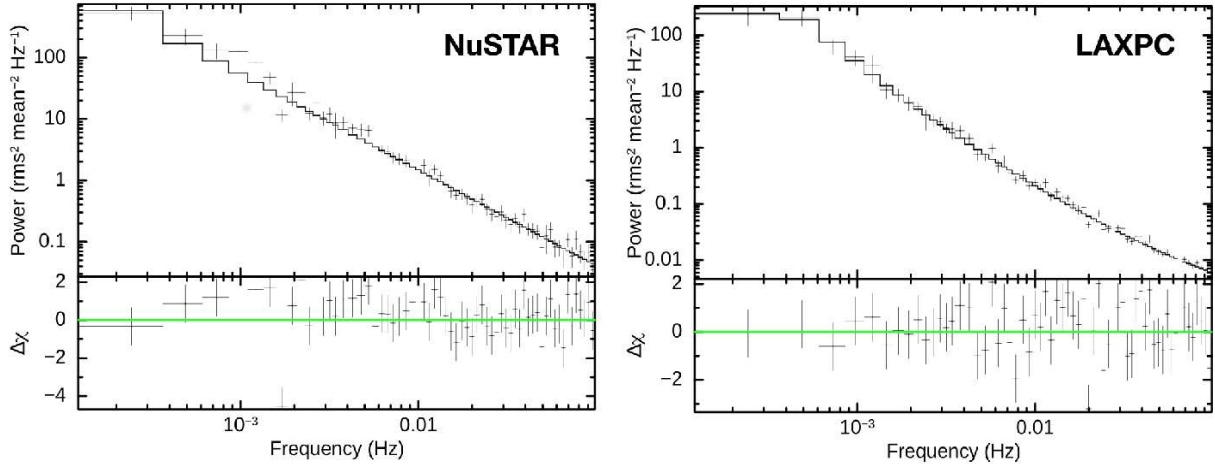}
\caption{Power density spectrum generated from 1s binned light-curves. Top panel: PDS obtained from the  3.0-79.0 keV combined light-curve of NuSTAR's FPMA and FPMB modules. Bottom panel: PDS obtained from 3.0-80.0 keV combined  light-curve of LAXPC10 and LAXPC20.} 
\label{fig:pds_lax_nu}
\end{figure*}

\begin{table*}
\caption{Best fit parameter values of  5.0-9.0 keV spectrum of NuSTAR's FPMA and FPMB fitted with TBabs*(powerlaw+gaus+gaus).}
\centering
\renewcommand{\arraystretch}{1.8}
\begin{tabular}{ccccccccc}
\hline
Photon Index & $N_H$ & $E(Fe K_{\alpha})$ & $\sigma(Fe K_{\alpha})$ & $Depth(Fe k_{\alpha})$ & $E(Ni K_{\alpha})$  & $\sigma(Ni K_{\alpha})$ & $Depth(Ni K_{\alpha})$ & $\chi^2/dof$ \\
- & $10^{22}$ $cm^{-2}$ & KeV & eV & - & KeV & keV & - & - \\ 
\hline 
 $1.47^{+0.03}_{-0.03}$ & $15.41^{+0.88}_{-0.95}$ & $6.36^{+0.01}_{-0.01}$ &  $41.25^{+39.89}_{-41.07}$ & $(9.33^{+0.82}_{-0.66})\times10^{-4}$ &  $7.56^{+0.05}_{-0.05}$ & $0.20^{+0.06}_{-0.06}$ &  $3.81^{+0.81}_{-0.75}\times10^{-4}$ & 78.51/88 \\

\hline
\end{tabular}
\label{tab:iron_line}%
\end{table*}

\subsection{Timing analysis}
We have used 131180s of stare time of AstroSat observation of 4U 1700-37 to achieve 48.62 ks effective exposure for our analysis. LAXPC has an excellent resolution of 10 $\mu$s, 
which is unique when combined with a large effective area of 6500 cm$^2$ \citep{2017JApA...38...30A, 2017ApJS_antia_LAXPC, 2016ExA....42..249R} in probing coherent or quasi periodicity. Visible gaps in the 15-50 keV LAXPC light curve pertain to source occultation and south atlantic anomaly (SAA) passages (Figure \ref{fig:all_lc}(c)). From the figure we can also see that the LAXPC and BAT light-curves follow a similar trend.  A combined light-curve of LAXPC10 and LAXPC20, with 1s bin size has been generated using \texttt{laxpc\_make\_lightcurve} module of \texttt{LAXPCsoftware}.

To look for previously reported periodicity from the source with the LAXPC data, the lightcurve has been extracted from all layers and for 3.0-80.0 keV energy range. The source had been observed in a bright state with a total count rate of 1043.7$\pm$0.15 $s^-1$ in the 3-80 keV band (flux $\sim$ $9.9\times10^{-9}$ $ergs/cm^2/s$,  as computed using the \texttt{flux} command after the spectral fitting of LAXPC data in XSPEC). The PDS have been generated from the combined 1s binned lightcurve using the ftool \texttt{powspec} of HEASOFT. The light
curve was segmented into stretches of 4096 bins per interval. Individual PDS of these intervals were averaged to obtain a final PDS which is shown in the bottom panel of Figure \ref{fig:pds_lax_nu}. To subtract the Poissonian noise, the PDS were normalized such that their integral gives the squared rms fractional variability normalized to units of (rms/mean)$^{2}$ Hz$^{-1}$. Further, to investigate periodicity/quasi-periodicity in 1-1000 Hz range, a combined light curve binned at 0.5 ms was obtained from LAXPC 10 and 20. This light curve was segmented in stretches of 2048 bins. White noise subtracted PDS were then generated from individual segments and averaged to obtain the final PDS. There is no detection of coherent pulsation or quasi-periodicity in the range 1-1000 Hz from the LAXPC data. 

Following the same procedure as above we have generated a PDS from a 1s binned combined lightcurve of FPMA and FPMB, for the energy range 3.0-79.0 keV  (top panel, Figure \ref{fig:pds_lax_nu}). To combine the FPMA and FPMB light-curves, we have used the ftool \texttt{fmerge}. The source is comparatively less bright during the NuSTAR observation. In 3.0-79.0 keV energy range the source count rate is found to be 71.3$\pm$0.07 counts/s. \par

 We have observed that the PDS obtained from the 1s binned LAXPC lightcurve can be described by the sum of a powerlaw and a zero centred Lorentzian model. We have ignored two spikes in the 
 data points corresponding to frequencies 0.027 and 0.065 Hz to obtain a reduced $\chi^{2}$ of 
 1.15 ($\chi^2/dof$= 64.82/56). The PDS of NuSTAR is well described with a powerlaw model and the reduced $\chi^{2}$ is found to be 1.19 ($\chi^2/dof$=72.80/61). We do not find the presence of any coherent pulsation or significant quasi-periodicity in the source. We also failed to detect the previously reported QPOs at 6.5 mHz \citep{1999A&A...349..873R} and $\sim$20 mHz \citep{2015MNRAS.448..620J}, in either the ASTROSAT or the NuSTAR data. \par

We have evaluated the orbital phases during these observations in the following manner. The BAT all sky monitor (ASM) lightcurve between 57000-58500 MJD has been barycenter corrected using the \texttt{earth2sun} package available in HEASOFT. Using the ftool \texttt{efsearch} on it, the orbital period is found to be $294764.55\pm68.78$ s ($\sim$3.411627 days), which is consistent with the result published earlier \citep{2016MNRAS.461..816I}. For NuSTAR and XRT data \texttt{barycorr} has been used to apply the barycentric correction and for LAXPC data the standard ASTROSAT tool \texttt{as1bary} has been used for the same. We compute orbital phases using the above period and a reference epoch of MJD 57446.55. A resulting folded light-curve is shown in Figure \ref{fig:all_lc}(a).

\begin{figure}
\centering
\includegraphics[width=\columnwidth]{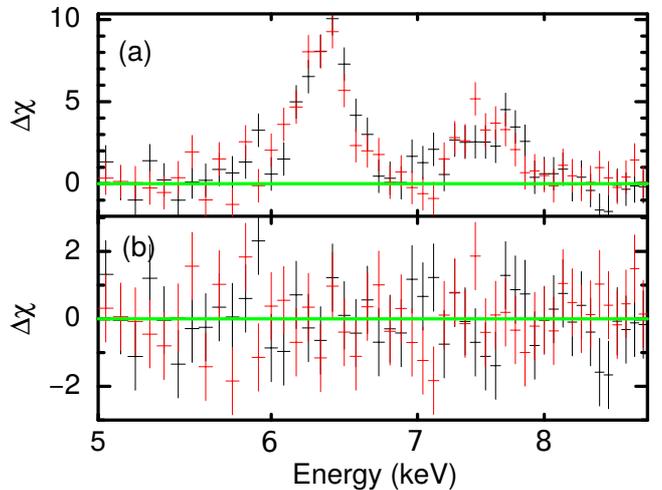}
\caption{The 5.0-9.0 keV spectrum of NuSTAR's FPMA and FPMB fitted with the model \texttt{cons*TBabs*(powerlaw+gaus+gaus)}. The two Gaussians are used to model the $\sim$6.4 keV and $\sim$7.6 keV lines. a) Plot of delchi after setting the norms of two Gaussian equal to zero  b) Plot of delchi for best fit model parameters.}
\end{figure}

\subsection{Spectral Analysis}
We have done a combined fit of NuSTAR's FPMA and FPMB data using the well known X-ray spectral fitting software XSPEC (V 12.10.1).
We have binned the data following the standard binning condition, i.e having a minimum 30 counts per bin and taking into account the instrument resolution (i.e bin size equal to 1/3 of FWHM) by combining 2, 
3, 5, 8, 16, 18, 30, 45, 76 channels for energies above 3, 10, 15, 20, 35, 45, 55, 64 and 70 keV respectively. This follows the prescription of \citet{2018A&A...620A.153F} upto 45 keV, and as we have more photons at higher energies, we have followed the standard condition above 45 keV.\par
The continuum of HMXBs is usually fitted with various empirical models like  high energy cutoff powerlaw \citep{1983ApJ.cutoffpl..270..711W}, Fermi Dirac cutoff powerlaw (FDCUT; \citet{1986LNP...255..198T}), NewHCUT  (a third order polynomial function with continuous 
derivatives; \citet{2000ApJ.newhcut..530..429B}), negative positive cutoff powerlaw or NPEX \citep{1999ApJ..npex.525..978M}. Some physical models like  Thermal Comptonization model (CompTT; \citet{1994ApJ...434..570T}), Thermal and bulk Comptonization of a seed blackbody-like spectrum (COMPTB; \citet{2008ApJ...680..602F}), thermal and bulk Comptonization for cylindrical accretion (COMPMAG; \cite{2012A&A...538A..67F}) have occasionally been used to describe the continuum. We have applied some of these models to describe the spectrum of 4U 1700-37. We have added a partial absorption \texttt{TBpcf} to describe the low energy part of the spectra. To describe a cyclotron feature we used a Gaussian absorption line model \texttt{gabs}.

To adjust factors related to cross-instrument calibration uncertainties, we have used a multiplicative model \texttt{constant} (cons) inbuilt in XSPEC along with the other models. This value of the constant is frozen at 1.0 for FPMA. For FPMB we have kept the parameter to be free. For all fits, the relative cross-instrument factors or the constant values for FPMB, are found to be nearly 1.03.

\subsubsection{\textbf{Probing emission lines of 4U 1700-37 using NuSTAR, SUZAKU and XMM-Newton}}
\textbf{\textit{NuSTAR:}} In the NuSTAR spectrum we find the presence of an iron 
K$_{\alpha}$ emission line at $6.36\pm0.02$ keV and 
another emission line at $7.56\pm0.05$ keV with a large 
width $\sim$0.2 keV. The effect of these on the high energy continuum model or the cyclotron features is negligible. Nonetheless, we have studied these emission lines by fitting the NuSTAR data in the energy 
range 5.0-9.0 keV, with a \texttt{powerlaw} with an 
absorption \texttt{TBabs} and two Gaussian emission 
(\texttt{gauss}) lines 
[\texttt{cons*TBabs*(powerlaw+gauss+gauss)}]. The best fit 
parameter values are given in Table \ref{tab:iron_line}.  
While fitting the complete spectra in the 5-75.0 keV band, 
whenever we are unable to constrain the low energy 
emission lines, we have fixed the line parameters to the
value quoted above.
Previously a Gaussian line at $\sim$6.5 keV was reported by (\citet{1999A&A...349..873R} from a BeppoSAX observation. \\

\noindent \textbf{\textit{XMM-Newton:}}  Earlier An XMM-Newton 
observation showed a prominent fluorescence Fe K$\alpha$ line at 
$\sim$6.4 keV, accompanied by the detection of a second K$\alpha$ line 
at a slightly higher energy $\sim$6.7 keV and a K$\beta$ line at $\sim$7.1
keV \citep{2005A&A...432..999V}.\par
We have analyzed the XMM-Newton observation (obs Id 0083280401; MJD 51960.74) of the source during the eclipse phase of Figure 1 of \citet{2005A&A...432..999V}. Among the XMM Newton detectors, namely three European photon imaging camera (EPIC); MOS1, MOS2, pn, and one Reflection Grating Spectrometer (RGS), only EPIC-MOS2 has observed the source in timing mode. Science Analysis Software (SAS), version 18.0, and the latest calibration files (as available on April 1, 2019) are used for the data reduction. The SAS task \texttt{emproc} has been used to generate a calibrated event file. The energy range of 0.3-10 keV is used as an energy filter\footnote{\url{http://xmm2.esac.esa.int/docs/documents/CAL-TN-0018.pdf}}. During this observation, XMM-Newton Epic-MOS2 data is free from pile up issues. The source region is selected as a circular region of radius 800 physical coordinates centred on the source, and the background spectrum is extracted from an annulus of width 200 physical coordinates surrounding the circular source extraction region. The spectrum of the source and the background regions are generated using SAS task \texttt{evselect}. The Redistribution Matrix File (RMF) and The Ancillary Response File (ARF) are created by using the SAS tasks \texttt{rmfgen} and \texttt{arfgen} respectively. Grouping of the spectra is done by the SAS task \texttt{specgroup}. The minimum count per group is set to 25, and the minimum energy width of each group is set to 3. \par
It has been 
found that along with the previously reported emission lines \citep{2005A&A...432..999V} a Gaussian 
emission line at 7.5 keV, improves the fit of the 0.3-10.0 keV XMM-Newton spectrum. The $\chi^2/dof$ is found to
be 660/605 without including the 7.5 keV line, and  653.3/602 after 
including the line.  The line is found with more than 2$\sigma$ 
confidence level, having a energy  $7.56^{+0.13}_{-0.10}$ keV and width 
$\sigma<$0.17 keV. As the detection level was less than 
3$\sigma$, the presence of the line could not be claimed from the 
XMM-Newton observation.\\

\noindent \textbf{\textit{SUZAKU:}} The iron lines at $\sim$6.7 keV and $\sim$7.1 keV were also found in a SUZAKU observation \citep{2015MNRAS.448..620J}, . 
 We have analysed publicly 
available SUZAKU data (processing date 27th May 2016) of 4U~1700-37, observed on 2006 September 13-14 (Obs ID: 401058010). The observation was performed in "XIS nominal" position with an effective exposure of $\sim$81 ~ks for XIS. The HEASOFT software package (version 6.26) and the
calibration database (CALDB) released  on 2015 October 05 (for XIS) 
are used for the data analysis. Unfiltered event files were processed by using \texttt{aepipeline} package of FTOOLS, along with standard screening criteria, to create cleaned XIS  event files. Following \citet{2015MNRAS.448..620J}, the photon pile-up for XIS data has been corrected by taking an annulus region with inner 
and outer radii of $30''$  and $180''$ respectively from the source position.
 Source spectra are extracted from 
the reprocessed XIS event data by selecting the above annulus region in XSELECT. Background spectra are generated by selecting circular 
regions away from the source. The response and effective area files for all the XIS detectors were generated by using the task \texttt{xisrmfgen} and \texttt{xissimarfgen} respectively. \par

To probe the low energy emission lines we have combined all the XIS spectra by using the ftool \texttt{addascaspec} as recommended by the XIS 
 team\footnote{\url{https://heasarc.gsfc.nasa.gov/docs/suzaku/analysis/abc/node9.html}}. This also combine all 
 \texttt{rmf}, \texttt{arf} and background spectrum files of different XIS units. We have used the combined spectra, background, \texttt{rmf} 
 and \texttt{arf} to look for the presence of low energy emission lines. Following \citet{2015MNRAS.448..620J} the combined spectra is binned by a factor of 6 from 0.8 to 10 keV. We find that in XIS
 spectra the inclusion of a 7.5 keV Gaussian line along with the previous reported lines in \citet{2015MNRAS.448..620J}, improves the fit ($\chi^2/dof$ = 445/365 without the line, and $\chi^2/dof$ = 411.8/362 with the 
 line). In the combined XIS spectra the 7.5 keV line has been detected with more than
 5$\sigma$ confidence level. The estimated line energy is $7.52^{+0.03}_{-0.03}$ keV and the width $\sigma<$0.06 keV. \par

\subsubsection{\textbf{Probing cyclotron absorption features in 4U 1700-37 with NuSTAR}}
We have used 3.0--75.0 keV NuSTAR data to probe any cyclotron line feature. 
 To describe the continuum of 4U 1700-37 we have applied the NPEX model 
\texttt{[cons*TBpcf*(powerlaw*npex+gaus+gaus)]}, following the previous work of \citet{2015MNRAS.448..620J}. The NPEX model has been 
created by adding two \texttt{cutoffpl} models with their cutoff energies tied to each other and keeping the photon index of one to be frozen at -2.0. For the best fit, the $\chi^2/dof$ is found to be 577.64/475. The fit shows some residuals in the overall spectrum. We
added a Gaussian absorption model around 39 keV (following the previous work of \citet{2015MNRAS.448..620J}),  but the best fit gives
the line energy as  $15.44^{+0.56}_{-0.53}$ keV with $\chi^2/dof$=487.32/472. The width and the
depth of the line are found to be $5.47^{+0.90}_{-0.78}$ keV and $1.29^{+0.51}_{-0.39}$
respectively. The chance probability of the line has been computed using the \texttt{ftest} task in XSPEC. The F-test with this absorption line gives an F value = 29.2 and a chance probability of $2.61\times10^{-17}$. If we add another Gaussian absorption line at 38.9 keV ( Energy value frozen) the best fit $\chi^2/dof$ is found to be 486.8/470. This indicates that the second absorption line is not required for the fit. 
If we use only one Gaussian absorption line and freeze the line energy at 38.9 keV then the width and the depth of the line are found 
to be  $4.54_{-0.87}^{+0.90}$ keV and $2.87_{-1.05}^{+1.24}$ respectively with a $\chi^2/dof$ = 
549.2/473. The \texttt{ftest} gives a chance 
probability of the 38.9 keV line to be $6.4\times10^{-6}$. So, with NPEX model we find two valid model combinations of the data. One, the presence of a Gaussian absorption line at $\sim$15 keV, two, the presence of a Gaussian absorption line at 38.9 keV. But, the presence of both lines together is not supported by the data. The 10.0-70.0 keV flux of the source is found to be
$(2.26\pm0.01)\times10^{-9}$ $ergs/cm^2/s$, much lower than the value
($5.6\pm0.3)\times10^{-9}$ $ergs/cm^2/s$, previously reported from SUZAKU data 
\citep{2015MNRAS.448..620J}.\par

We have observed that the $\sim$16 keV absorption line structure found with the NPEX model is not smooth but has a double peak. So, we added another absorption component having 
energy 

\begin{figure*}

\centering
\includegraphics[width=17cm, height=14.cm]{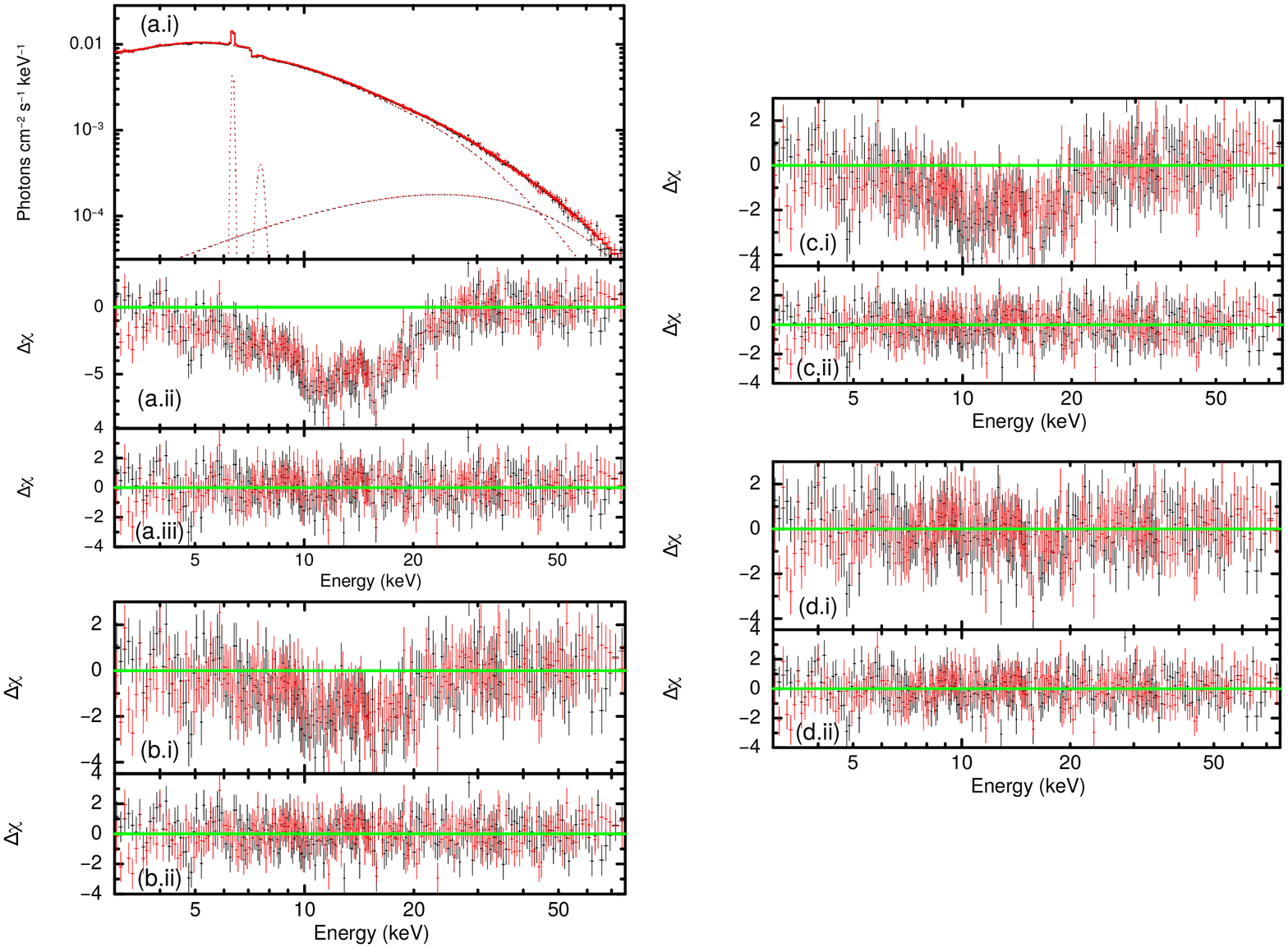}
\caption{Spectrum of 4U 1700-37 obtained from NuSTAR's FPMA (black) and FPMB 
(red), along with the best fit models; a) HCUT 
\texttt{[cons*TBpcf*(powerlaw*hcut+gaus+gaus)]}, b) NewHCUT 
\texttt{[cons*TBpcf*(newhcut*powerlaw+gaus+gaus)]}, c) NPEX 
\texttt{[cons*TBpcf*(npex+gaus+gaus)]} d) COMPMAG 
\texttt{[cons*TBpcf*(compmag+gaus+gaus)]}. The possible cyclotron absorption 
line is shown after fitting a gabs and then setting the depth to zero (a.ii, 
b.i, c.i, d.i, d.iii) and the corresponding bottom panels (a.iii, b.ii, c.ii, 
d.ii, d.iv) are the delchi plot for best spectral fitting.}
\label{1_cyc_line_TBpcf}
\end{figure*}

\noindent  $11.15^{+0.48}_{-0.23}$ keV and it marginally improves the fit ($\chi^2/dof$=$474.24/469$). The  width and depth of this second line is found to be $\sigma_2=0.96^{+0.65}_{-0.31}$ keV  and $d_{cyc2}=0.04^{+0.05}_{-0.02}$ respectively. Although the overall continuum is not affected after including the second absorption line, the parameters of the first absorption line get slightly modified. The energy, width, and depth of the modified first line are
found to be  $E'_{cyc1}=16.14^{+0.71}_{-0.58}$ keV, $\sigma'_{cyc1}=5.19^{+0.93}_{-0.87}$ keV and  $d'_{cyc1}=0.95^{+0.46}_{-0.37}$  respectively. All the best fit parameters are given in Table \ref{2nd_cyclotron_line}. The \texttt{ftest} gives the chance probability of the second line as $0.005$. \par

We have observed similar kind of residuals around 15.0-17.0 keV after fitting the continuum with a NewHCUT model.The model  
\texttt{cons*TBpcf(newhcut*powerlaw*gabs+gaus+gaus)} gives the $\chi^2/dof$ is to be $531.44/475$. With this continuum, if we include a 38.9 keV absorption line, the depth is found to be very low and almost no change occurs in $\chi^2$ value. If we freeze the width at the previous published value of 9.8 keV \citep{2015MNRAS.448..620J}, then the upper limit to the depth is found to be 0.11 ($\chi^2/dof$=531.45/474).
The fit is found to improve significantly ($\chi^2/dof$= $481.03/472$) when we allow the line energy to be free and the line energy is found to be $E_{cyc1}$=$15.82^{+1.11}_{-1.27}$ keV. The best fit parameter values are given in Table \ref{line_NuSTAR}. In this case also 
the absorption line is found to have a double peak and the inclusion of a second line improves the fit 
(table--\ref{2nd_cyclotron_line}).\par

To check the consistency of the absorption lines, we have also applied the HCUT model. The model \texttt{cons*TBpcf*(powerlaw*hcut+gaus+gaus)} 
also gives an acceptable fit ($\chi^{2}/dof$=516.65/475) but has some residuals around
15.0--17.0 keV.  Modelling the residuals with \texttt{gabs} 
gives a wide absorption line. The reason could be the sharp feature produced by the HCUT model at the cutoff energy \citep{1997ESASP.Kretschmar.382..141K, 1999A&A.Kreykenbohm..341..141K}.  In 
general practice a Gaussian
optical-depth profile is added to smooth the transition 
between the pure power law components and the cutoff power law \citep{2002ApJ...580..394C, 2013arXiv1309.5361F}. To implement this, we add a \texttt{gabs} component  tied to the
cutoff energy  $7.37^{+0.20}_{-0.29}$ keV. The width and depth 
of this \texttt{gabs} is found to be $0.54\pm0.14$ keV and 
$0.05\pm0.02$ respectively. As the cutoff energy is close to 
the Ni $K_{\alpha}$ line energy, we have kept the energy and 
the width of the Ni $K_{\alpha}$ line frozen to the values mentioned in 
table \ref{tab:iron_line}.
This gives the best fit ($\chi^2/dof$=$471.78/472$) cyclotron line energy to be  $15.45^{+0.94}_{-0.90}$ keV
having width and depth $5.12^{+1.92}_{-1.40}$ keV and  
$0.49^{+0.50}_{-0.25}$ respectively. With this continuum if 
we fix the energy of absorption line to 38.9 keV, 
then the best fit ($\chi^2/dof$=516.64/470) gives the depth to 
be very low. If we keep the width to be frozen to the value 
7.3 keV \citep{2015MNRAS.448..620J}, then the upper limit to the depth of the line is found to be 0.12.\par

In this case also, the $\sim$16 keV absorption line 
is found to have a double peak, and the inclusion of another
Gaussian absorption line at $10.83^{+0.69}_{-0.54}$ keV 
slightly improves the fit ($\chi^2/dof$ = $463.74/469$). The 
width and the depth of the second line are found to be 
$\sigma_2$= $1.28^{+0.81}_{-0.57}$ keV and  
$d_{cyc2}$=$0.06^{+0.06}_{-0.04}$ respectively  
(Table-\ref{2nd_cyclotron_line}).\par

We then use a physical model COMPMAG to describe the continuum of the spectrum. This model assumes a cylindrical polar cap accretion on the NS due to the 
presence of the magnetic field. In this model, it is considered that soft photons with temperature $T_{bb}$ are upscattered by the in-falling plasma having a 
temperature $T_e$. This model gives the best fit when we set the \texttt{betaflag} to 2.0 which makes the model independent of the parameters 
\texttt{$\beta_0$} and \texttt{$\eta_0$}. In this case, we are not able to 
compute the lower limit of the width of the $\sim$7.6 keV emission line. So we mention the upper limits. We get an acceptable fit with the model \texttt{cons*TBpcf*(compmag+gaus+gaus)} ($\chi^2/dof$=$491.15/473$). 

 When we add a \texttt{gabs} 
around 38.9 keV, it does not improve the fit, and its depth and width could not be computed simultaneously. On the other hand, if we include one 
Gaussian absorption component at $17.27^{+1.14}_{-0.96}$ keV, the fit improves a little ($\chi^2/dof$=$477.05/470$). The width and the depth of the line are found to be 
$\sigma_1=2.28^{+1.64}_{-0.66}$ keV and

 \begin{landscape}
\begin{table}
\fontsize{11}{3}
\caption{Best fit model parameters obtained from  NuSTAR's FPMA and FPMB (3.0-75.0 keV) data. The combination of models are given in the caption of Figure \ref{1_cyc_line_TBpcf}. The $\dagger$ symbol indicates that a Gaussian absorption component (\texttt{gabs}) has been added with the continuum model.} 
\renewcommand{\arraystretch}{2.0}
\begin{tabular}{p{2.9cm}p{2.1cm}p{2.1cm}p{2.1cm}p{2.1cm}p{2.1cm}p{2.1cm}p{2.1cm}p{2.1cm}}
\hline
Parameter & NPEX  & NPEX$\dagger$ &  NewHCUT & NewHCUT$\dagger$ & COMPMAG  & COMPMAG$\dagger$  & HighECut & HighECut$\dagger$ \\
\hline \hline
$Photon Index$ &                      $0.79^{+0.02}_{-0.02}$ &  $0.81^{+0.05}_{-0.04}$ &  $1.32^{+0.02}_{-0.03}$ &  $1.28^{+0.04}_{-0.04}$ & - & - &  $1.44^{+0.04}_{-0.04}$ &  $1.32^{+0.02}_{-0.03}$  \\
$N_{H}\;(10^{22}\;cm^{-2})$ &      $19.58^{+0.89}_{-0.91}$ & $19.83^{+0.87}_{-0.89}$ &   $18.77^{+0.82}_{-0.84}$ & $19.40^{+0.84}_{-0.85}$ & $20.82^{+0.66}_{-1.07}$ &  $20.75^{+0.90}_{-0.40}$ & $18.87^{+0.87}_{-0.84}$ &   $19.83^{+0.93}_{-0.90}$ \\
$f$ &                              $0.79^{+0.01}_{-0.01}$ &  $0.78^{+0.01}_{-0.01}$ &    $0.82^{+0.01}_{-0.01}$ &  $0.80^{+0.01}_{-0.01}$ & $0.68^{+0.01}_{-0.01}$ &  $0.69^{+0.01}_{-0.01}$ &  $0.81^{+0.01}_{-0.01}$ &  $0.80^{+0.01}_{-0.01}$  \\
$E_{cut}$ (keV) &                   -  & - &   $7.38^{+0.26}_{-0.26}$ &  $7.58^{+0.56}_{-0.38}$ & - & -& $7.39^{+0.21}_{-0.23}$ &  $7.37^{+0.20}_{-0.29}$ \\
$E_{fold}$ (keV) &                   $9.63^{+0.21}_{-0.19}$ & $11.91^{+1.13}_{-0.82}$ &  $24.69^{+0.70}_{-0.66}$ &  $24.19^{+1.20}_{-1.16}$ & - & -&   $24.77^{+0.66}_{-0.67}$ &   $23.38^{+0.81}_{-0.79}$  \\
$kT_{BB}$ (keV) &                   - & - & -& -&  $1.35^{+0.02}_{-0.03}$ &  $1.32^{+0.03}_{-0.03}$ &  -&  - \\
$kT_{e}$ (keV) &                    - & - & -& -&  $4.28^{+0.00}_{-0.52}$ &   $3.88^{+0.14}_{-0.28}$ & -&  -\\
$\tau$ &                            - & - & -& -&  $0.38^{+0.03}_{-0.05}$ & $0.39^{+0.10}_{-0.05}$ &  -&    -\\
$r_{0}$ &                           - & - & -& -&  $0.23^{+0.04}_{-0.02}$ &  $0.23^{+0.02}_{-0.04}$ &  -&    -\\
$A$ &                               - & - & -& -& $0.63^{+0.02}_{-0.08}$ & $0.61^{+0.33}_{-0.09}$ &   -&    -\\
$norm$ &                              - & - & -& -& $106.02^{+6.62}_{-6.60}$ &  $111.30^{+12.47}_{-6.82}$ &   -& - \\
$E_{cyc}$ (keV) &                  - &  $15.44^{+0.56}_{-0.53}$ & -&  $15.82^{+1.11}_{-1.27}$ & - & $17.27^{+1.14}_{-0.96}$ &  -&   $15.45^{+0.94}_{-0.90}$ \\
$\sigma_{cyc}$ (keV) &             - &  $5.47^{+0.90}_{-0.78}$ & -&  $4.73^{+1.97}_{-1.41}$ & -& $2.28^{+1.64}_{-0.66}$ &  -&  $5.12^{+1.92}_{-1.40}$\\
$d_{cyc}$ &                        - &   $1.29^{+0.51}_{-0.39}$ &  - &  $0.42^{+0.48}_{-0.20}$ & -& $0.10^{+0.22}_{-0.07}$ &  - &    $0.49^{+0.50}_{-0.25}$ \\
$significance$ &                       - &  $5.3$               & -  &        $4.1$              & - &         $2.8$         & -  &        $3.0$    \\
$Chance\;Pro$ &                - & $2.61\times10^{-17}$ & - & $3.31\times10^{-10}$ & - & $0.003$ & - & $2.54\times10^{-9}$ \\
$E(K_{\alpha})$ (keV) &              $6.35^{+0.01}_{-0.01}$ &  $6.35^{+0.01}_{-0.01}$ & $6.35^{+0.01}_{-0.01}$ & $6.35^{+0.01}_{-0.01}$ & $6.35^{+0.01}_{-0.01}$ & $6.35^{+0.01}_{-0.01}$ & $6.35^{+0.01}_{-0.01}$ & $6.35^{+0.01}_{-0.01}$ \\
$\sigma(K_{\alpha})$ (keV) &         $0.07^{+0.03}_{-0.03}$  & $0.05^{+0.04}_{-0.05}$ &  $0.07^{+0.03}_{-0.05}$ &  $0.05^{+0.04}_{-0.05}$ & $0.02^{+0.06}_{-0.01}$  & $0.05^{+0.03}_{-0.04}$  &  $0.05^{+0.03}_{-0.05}$ &  $0.05^{+0.03}_{-0.05}$  \\
$E(K_{\beta})$ (keV) &              $7.58^{+0.06}_{-0.06}$ & $7.58^{+0.06}_{-0.06}$ &   $7.57^{+0.06}_{-0.06}$ &  $7.57^{+0.06}_{-0.06}$ & $7.58^{+0.10}_{-0.10}$ &  $7.57^{+0.10}_{-0.10}$ & $7.56$ &  $7.56$ \\
$\sigma(K_{\beta})$ (keV) &          $0.22^{+0.10}_{-0.08}$ &  $0.16^{+0.08}_{-0.09}$  & $0.22^{+0.08}_{-0.07}$ & $0.15^{+0.07}_{-0.08}$ & $<0.17$ & $<0.18$ & $0.2$ & $0.2$ \\
\hline
$\chi^2/dof$ &                       $577.64/475$ &  $487.32/472$ & $531.44/475$ & $481.03/472$ & $491.15/473$ & $477.05/470$ & $516.65/475$ & $471.78/472$ \\
$\chi^2_{red}$ &                    $1.22$ & $1.03$ & $1.12$ & $1.02$ & $1.04$ & $1.02$ & $1.09$ & $1.00$\\
\hline \hline
\end{tabular}
\label{line_NuSTAR}
\end{table}
\end{landscape}

\noindent $d_{cyc1}=0.10^{+0.22}_{-0.07}$ respectively. With this continuum model, we have not noticed any distortion in the absorption line.
We find the soft photon temperature $kT_{BB}\sim$1.32 keV and the plasma temperature $kT_e\sim$3.88 keV. The albedo $A$ is found to be $\sim$0.6, which is consistent with a reflection from the neutron star surface (0<$A$<1). The parameter $r_0$ is found to be nearly 0.23 (~0.8 km) which is considered to be the radius of the accretion column, in units of NS Schwarzschild radius. The distance of the source ($\sim$1.9 kpc) and the normalisation constant $(R_{km}/d_{10kpc})^2$ $\sim110$ give the soft photon source radius to be $\sim$1.9 km.\par

We have also conducted a combined fitting of ASTROSAT SXT, LAXPC and CZTI spectrum. The uncertainties in LAXPC response and the poor SNR of the source spectrum in LAXPC and CZTI, do not allow us to probe the presence of a cyclotron line using ASTROSAT data.

\begin{table}
\fontsize{10}{2}
\caption{Best fit model parameters obtained from a combined fit of  NuSTAR's FPMA and FPMB (3.0-75.0 keV) data. The combination of models are given in the caption of Figure \ref{1_cyc_line_TBpcf}. The $\dagger$ symbol indicates that two additional Gaussian absorption components (\texttt{gabs}) have been added to the continuum model.} 
\renewcommand{\arraystretch}{2.0}
\begin{tabular}{p{2.2cm}p{1.5cm}p{1.5cm}p{1.5cm}}
\hline
Parameter & NPEX  & NewHCUT & HCUT \\
\hline \hline
Photon Index &                      $0.83^{+0.07}_{-0.05}$ & $1.29^{+0.13}_{-0.04}$ & $1.30^{+0.03}_{-0.03}$   \\ 
$N_{H}\;(10^{22}\;cm^{-2})$ &     $19.84^{+0.84}_{-0.89}$ & $19.46^{+0.83}_{-0.86}$ & $19.81^{+0.97}_{-0.94}$  \\
$f$ &                            $0.78^{+0.01}_{-0.01}$ & $0.80^{+0.01}_{-0.01}$ &  $0.80^{+0.01}_{-0.01}$ \\
$E_{cut}$ (keV) &                   -                       & $7.32^{+2.89}_{-0.60}$ & $7.48^{+0.19}_{-0.28}$ \\
$E_{fold}$ (keV) &                  $12.41^{+1.47}_{-1.07}$ & $23.61^{+0.98}_{-0.95}$ & $23.55^{+0.80}_{-0.80}$ \\
$E'_{cyc1}$ (keV) &                  $16.14^{+0.71}_{-0.58}$ & $16.55^{+0.79}_{-0.85}$ & $16.49^{+0.82}_{-0.88}$ \\
$\sigma'_{cyc1}$ (keV) &              $5.19^{+0.93}_{-0.87}$ & $3.03^{+2.19}_{-0.95}$ & $3.37^{+1.76}_{-1.13}$  \\
$d'_{cyc1}$ &                         $1.18^{+0.47}_{-0.36}$ & $0.24^{+0.38}_{-0.09}$ & $0.31^{+0.28}_{-0.13}$ \\
$significance 1$ &                       $5.3$ & $4.0$ & $3.3$ \\
$E_{cyc2}$ (keV) &                  $11.15^{+0.48}_{-0.23}$ &  $11.06^{+0.72}_{-0.68}$ & $10.83^{+0.69}_{-0.54}$ \\
$\sigma_{cyc2}$ (keV) &              $0.96^{+0.65}_{-0.31}$ & $1.07^{+0.98}_{-0.37}$  & $1.28^{+0.81}_{-0.57}$ \\
$d_{cyc2}$ &                         $0.04^{+0.05}_{-0.02}$ & $0.04^{+0.09}_{-0.03}$ & $0.06^{+0.06}_{-0.04}$\\
$significance2$ &                       $2.8$ & $2.0$ & $2.2$ \\
$E(K_{\alpha})$ (keV) &              $6.35^{+0.01}_{-0.01}$ & $6.35^{+0.01}_{-0.01}$ & $6.36^{+0.01}_{-0.01}$ \\
$\sigma(K_{\alpha})$ (keV) &         $0.05^{+0.04}_{-0.04}$ & $0.04^{+0.04}_{-0.04}$ & $0.03^{+0.04}_{-0.03}$\\
$E(K_{\beta})$ (keV) &               $7.58^{+0.06}_{-0.06}$ & $7.57^{+0.06}_{-0.06}$ & $7.56$  \\
$\sigma(K_{\beta})$ (keV) &          $0.13^{+0.08}_{-0.13}$ & $0.14^{+0.07}_{-0.10}$ &  $0.2$ \\
\hline
$\chi^2/dof$ &                       $474.24/469$ & $473.36/469$ & $463.74/469$\\
$\chi^2_{red}$ &                    $1.01$ & $1.00$ & $0.99$\\
\hline \hline
\end{tabular}
\label{2nd_cyclotron_line}
\end{table}

\section{Discussion and conclusions}
The source 4U 1700-37 is one of the HMXB pulsar candidates where the X-ray pulsations remain unconfirmed. The debate persists since the discovery 
of the source regarding the existence of a 97 min or 50s or 67.4s coherent periodicity. In this work, we have used high-resolution LAXPC data during 
the orbital phase 0.35-0.75. The source was bright during in this orbital phase with an observed flux of $(9.89_{-0.04}^{+0.01})\times10^{-9}$ 
$ergs/cm^2/s$. With the 3.0-80.0 keV LAXPC lightcurve, we have not found any sign of coherent or quasi-periodic oscillation in the PDS in the 
frequency range from 0.1 mHz to 10$^3$ Hz. A NuSTAR observation of this source had been carried out spanning the orbital phase 0.55 
to 0.80. During this phase the source was at a lower flux ($(2.26\pm0.01)\times10^{-09}$ $ergs/cm^2/s$) state.  The PDS of the NuSTAR observation 
was devoid of any coherent pulsation or QPO in the 3.0-79.0 keV energy range. The pulsation in X-ray pulsars originates due to a misalignment 
between the spin axis and the magnetic axis, causing the X-ray emitting hot spot to rotate in and out of the line of sight. The lack of observed 
pulsations in high magnetic field pulsars ($\sim$10$^{12}$ Gauss) is possible if the angle between the magnetic axis and the rotational axis 
of the pulsar is small.   
\par

In our spectral analysis of the NuSTAR data we find the 
presence of an emission line at $\sim$7.56 keV which has 
never been reported in this source. We detect this feature 
at more than 7$\sigma$ confidence. The line is identified 
as a rare Ni $K\alpha$ line which has occasionally been reported  in 
X-ray sources such as IGR J16318-4848 (galactic X-ray 
binary) \citep{2003A&A...411L.427W}, and in some AGNs 
\citep{2005MNRAS.360.1123P}. In a wind fed 
accretion powered X-ray pulsar, having neutral Ni in 
abundance, the stellar wind can ionize  and result in the
formation of a Ni K$\alpha$ ($\sim$7.5 keV) emission line \citep{1998arXiv:astro-ph/9810018}. 
\citet{2003A&A...411L.427W} suggested that the X-ray 
spectrum of an accreting high mass X-ray binary 
like IGR J16318-4848 features strong photo-electric 
absorption due to stellar wind. This leads to the 
detection of the Ni K$\alpha$ (7.5 keV) line.

We have found that the continuum can be described by the
empirical models NPEX, NewHCUT, HCUT and also by a physical model COMPMAG. 
We find the absorbing column density to the source to be high, $\sim$20$\times$10$^{22}$ cm$^{-2}$. This is consistent with the earlier studies of the source as mentioned in section 1. With the empirical continuum models we find strong residuals around 15-16 keV ($E_{cyc1})$, which can be modelled with a Gaussian absorption line and this improves the fit significantly. All reported lines with the empirical models are at least $3\sigma$ significant and the maximum chance probability of the line is found to be 2.27$\times10^{-9}$ (HCUT model). It has also been observed that the line shape deviates from a simple Gaussian profile. We found a marginal 
improvement in chi-square value after including another absorption line with an energy $\sim$11 keV ($E_{cyc2}$).
For all three empirical models the 
parameters of both the lines are found to be consistent. On the other hand if we describe the continuum with a
COMPMAG model, we find some residual around 17.0 keV. Including a Gaussian absorption line 
improves the fit. The second cyclotron line could not be discerned with this continuum model. It has 
also been noticed that with the COMPMAG model it is difficult to constrain the width of weak features 
like
the 7.6 keV emission line.\par

It has also been found that with different continuum models, a Gaussian absorption line at 38.9 keV previously reported from this source \citep{2015MNRAS.448..620J}, does not improve the fit significantly. Apart from NPEX none of these models allow us to compute the width and the depth of the 
38.9 keV line. So, with this data, we do not find any significant hint of the presence of a line at 38.9 keV. We have computed the upper limit to the line using the previously reported width \citep{2015MNRAS.448..620J}. One of the possible reasons for not finding the 38.9 keV line could be the difference in the flux state between these two observations. The source flux during the SUZAKU observation was nearly twice that during the NuSTAR observation. Another possibility could
be that the previously reported line was an artifact of an inadequately modeled continuum. \citet{2015MNRAS.448..620J}
used the NPEX continuum and concluded that the cyclotron line feature has a huge width ($\sim$19 keV), which is difficult to justify with any physical model. It has also been shown by \citet{2013A&A...551A...6M} that the NPEX model sometimes results in very broad cyclotron lines and line parameters cannot be blindly trusted. The first reported
cyclotron line by \citet{1999A&A...349..873R} was around ~37 keV. They too have suggested that the feature may
arise due to incorrect modeling of the continuum or a purely instrumental effect. \par

If the $\sim$16 keV cyclotron line is true, then it suggests that the magnetic field strength of the line forming region is nearly $2\times10^{12}$ Gauss. If the line is distorted and consists of two absorption profiles having energies around 11 keV and 16 keV, then the actual magnetic field strength may be  around
$1.4\times10^{12}$ Gauss. The presence of the 16 keV line along with the 11 keV residual could be described as a 
distorted line like that in Cep X-4 \citep{2015ApJ...806L..24F} or it may consist of two lines with anharmonic ratio. The ratio between the line energies in this case is $\sim$1.45, too small to be claimed as harmonics. In other sources, deviations from exact integral ratio have been observed, but the deviation is found to be much smaller (e.g., 4U 0115+63 \citet{1999ApJ...521L..49H}). Cep X-4 \citep{2015ApJ...806L..24F} is the only source where the line energy ratio is nearly 1.56.
Two CRSFs of such different energies may indicate a strong deviation from a simple dipole magnetic field, as might be expected due to the formation of an accretion mound \citep{2012MNRAS.420..720M}. The formation of accretion mound causes a redistribution of the magnetic field inside the accretion mound. The central field lines are dragged towards the edge, to balance the gravitational pressure of the mound by the magnetic field pressure. This reduces or keeps unaffected (depending on the mound profile) the magnetic field intensity at the center of the mound While the magnetic field intensity towards the edge of the mound increases. The integrated emission from different parts
of the mound causes broadening and asymmetry in the observed cyclotron line profile. An asymmetric line profile may also arise from simultaneous viewing of different parts of an accretion column with different field strengths. If the magnetic field intensity is not symmetric with respect to the column axis, the phase averaged cyclotron line will have an asymmetric shape, due to different magnetic field intensities in different phases.
Deviations from a smooth cyclotron line could also be attributed to the presence of strong emission wings, which is predicted from simulations \citep{2007A&A...472..353S}.\par 

Our spectral study of 4U 1700-37 with NuSTAR finds the presence of a rare $\sim$7.6 keV Ni K$\alpha$ line, which is being reported for first time in this source. Our study also finds the hint of a previously unreported cyclotron feature around 16 keV. In certain continuum models, this line feature deviates in shape from a simple Gaussian, suggesting the presence of a complex magnetic field structure. However this cannot yet be conclusively established. Our timing study with ASTROSAT and NuSTAR show no evidence of any periodic or quasi-periodic oscillation. Further observations at different flux states of the source would be important for a more detailed exploration of its spectral and temporal properties.

\section*{Acknowledgements}
 S. Bala and J. Roy want to thank Prof. Ranjeev Mishra and Prof. Gulab Chand Dewangan and Kabir Chakravarti for their useful suggestions and discussions. Authors would like to thank the annonymous referee for valuable comments and suggestion which improved the manuscript substantially. The research work is funded by the University Grants Commission (UGC)  and the Indian Space Research Organisation (ISRO). The authors want to thank Inter University Center for Astronomy and Astrophysics (IUCAA) for the research facilities. This work  made use of data from the NuSTAR mission, a
project led by the California Institute of Technology, managed by the Jet Propulsion Laboratory, and funded by the
National Aeronautics and Space Administration. This research has made use of the NuSTAR Data Analysis Software
(NuSTARDAS) jointly developed by the ASI Science Data Center (ASDC, Italy) and the California Institute of
Technology (USA). This publication uses the data from the AstroSat mission of the Indian Space Research Organisation (ISRO), archived at the Indian Space Science DataCentre (ISSDC). This work has used the data from the LAXPC developed at TIFR, Mumbai, and the LAXPC POC at TIFR is thanked for verifying and releasing the data via the ISSDC data archive and providing the necessary software tools. This research has made use of data, software and/or web tools obtained from the High Energy Astrophysics Science Archive Research Center (HEASARC), a service of the Astrophysics Science Division at NASA/GSFC and of the Smithsonian Astrophysical Observatory's High Energy Astrophysics Division. 
 We want to thank the instrument teams of SUZAKU and XMM-Newton. 



\bibliographystyle{mnras}
\bibliography{main} 


 $11.12^{+1.07}_{-0.67}$


\bsp	
\label{lastpage}
\end{document}